\documentclass[fleqn,10pt]{wlscirep}
\usepackage[utf8]{inputenc}
\usepackage[T1]{fontenc}
\usepackage{amssymb}
\usepackage{amsmath}
\usepackage{blkarray}
\usepackage{xcolor}
\usepackage[normalem]{ulem}

\definecolor{forestgreen}{rgb}{0.33,0.61,0.34}
\definecolor{darkteal}{HTML}{008080}

\title{Detecting and forecasting tipping points from sample variance alone}

\author[a,b,c*]{Naoki Masuda} 
\affil[a]{Gilbert S.\,Omenn Department of Computational Medicine and Bioinformatics, University of Michigan, Ann Arbor, MI 48109, USA}
\affil[b]{Department of Mathematics, University of Michigan, Ann Arbor, MI 48109, USA}
\affil[c]{Center for Computational Social Science, Kobe University, Kobe, Hyogo 657-8501, Japan}
\affil[*]{naokimas@gmail.com}

\begin{abstract}

Anticipating tipping points in complex systems is a fundamental challenge across domains. Traditional early warning signals (EWSs) based on critical slowing down, such as increasing sample variance, are widely used, but their ability to reliably indicate imminent bifurcations and forecast their timing remains limited. Here, we introduce TIPMOC (TIpping via Power‑law fits and MOdel Comparison), a parametric framework designed to statistically detect the approach of a bifurcation and estimate its future location using only the sample variance. TIPMOC exploits the mathematical property that variance diverges with a characteristic power-law form near codimension-one bifurcations. By sequentially monitoring system variance as a control parameter changes, TIPMOC statistically adjudicates between linear and power-law divergence at each step. When evidence favors power-law divergence, TIPMOC forecasts the impending tipping point and estimates its position; otherwise, it avoids false positives. Through numerical simulations, we demonstrate TIPMOC’s robustness and accuracy in both detection and timing prediction across different types of dynamics and bifurcation, whereas the accuracy of timing prediction is limited. TIPMOC shows low false positive rates and performs well even with uneven sampling and colored noise. This method thus enhances the interpretability and practical utility of classical EWSs, serving as both a transparent add-on and a stand-alone statistical tool for forecasting regime shifts in diverse complex systems.

\end{abstract}

\begin{document}

\flushbottom
\maketitle

\section*{Significance statement}

Climate, ecosystems, and other complex systems can abruptly change state when a slowly varying condition crosses a critical threshold (a tipping point). Yet widely used warning signs such as rising sample variance often give ambiguous alarms and rarely predict when the threshold will be reached. We propose TIPMOC, a statistical method that monitors variance alone and uses model comparison to distinguish the characteristic accelerating rise expected near a critical threshold from ordinary trends, enabling both detection of an approaching tipping point and an estimate of its location. Simulations across multiple bifurcation types show high detection rates, low false positives, and robustness to uneven sampling and colored noise, providing a transparent add-on to classical early-warning analyses.

\section*{Introduction}

Nonlinear dynamical systems provide a powerful language with which to describe, predict, and control various real-world phenomena. They often experience qualitative changes, which may be a discontinuous regime shift, often referred to as tipping events, or a continuous onset of qualitatively different dynamics, as the internal state of the system or its external environment changes. Such qualitative changes, including tipping events, are implicated in, e.g., species extinctions in ecosystems~\cite{Scheffer2009Nature, Scheffer2015AnnuRevEcolEvolSyst}, deforestation \cite{LiuKumar2019NatClimateChange,Wunderling2022PNAS}, other forms of climate change \cite{Boers2021NatClimateChange, Lenton2024NatCommun}, epidemic outbreaks~\cite{Southall2021JRSocInterface, Delecroix2023PlosGlobalPublicHealth}, progression of mental disorders~\cite{Vandeleemput2014PNAS, Dablander2023PsycholMethods}, and that of
%
%
somatic diseases~\cite{LiuChang2019NatSciRev,Aihara2022Gene}.
Although there are important exceptions \cite{Boettiger2013TheorEcol, Dakos2015PhilTransRSocB}, these qualitative shifts occur via a bifurcation of the underlying dynamical system in many cases.
Critical slowing down, which dictates that the relaxation time of the dynamics increases near an equilibrium as the bifurcation point is approached, provides an opportunity to construct and deploy early warning signals (EWSs) that aim to anticipate an impending bifurcation.
Various scalar EWSs based on critical slowing down, such as the sample variance, lagged autocorrelation, and skewness, have been proposed~\cite{Scheffer2009Nature, Dakos2012PlosOne,Scheffer2012Science, Boettiger2013TheorEcol, Dakos2015PhilTransRSocB,Southall2021JRSocInterface, George2023PhysicaScripta, Rietkerk2025NatClimateChange}.
Despite more recent advancements in EWSs using model comparison \cite{Boettiger2012JRSocInterface, Perretti2012EcolAppl, Hessler2022NewJPhys, Hessler2022PnasNexus, Hessler2025NatCommun} and machine learning \cite{Kong2021PhysRevResearch, Patel2023Chaos, Huang2024NatMachineIntel, Panahi2024PhysRevResearch, LiuZhang2024PhysRevX}, these conventional scalar EWSs remain widely used.

However, tools for assessing performance of scalar EWSs have been limited. A commonly used performance measure is Kendall's $\tau$ \cite{Dakos2012PlosOne, ChenGhadami2022RSocOpenSci}. 
Let us suppose that a dynamical system generating observed data depends on a slowly varying control parameter $u$ and that an EWS, denoted by $s(u)$, increases as $u$ approaches the bifurcation value. Let $\tau$ denote Kendall's rank correlation between $u$ and $s(u)$. A large $\tau$ is typically interpreted to correspond to a better EWS performance. Nonetheless,
Kendall's $\tau$ has at least three limitations.
First, one does not know which value of $\tau$ is large enough to infer an upcoming bifurcation.
Second, related to the first limitation, $\tau$ can be close to the maximum possible value, i.e., $1$, even when no bifurcation is imminent. For example, a simple noise process with a gradually increasing noise amplitude without other features can produce a large $\tau$ (we show this example in Table~\ref{tab:detection-summary}), which is a false positive. Third, a large value of $\tau$, or $\tau$ exceeding a predetermined threshold, does not quantify the remaining distance to the bifurcation in $u$.

Some modern EWS methods address the limitations of Kendall's $\tau$ listed above. First, methods based on model comparison compare goodness of fit of a model of stochastic dynamics exhibiting a bifurcation and a non-bifurcating alternative, and judge which model fits the observed data better. Therefore, by design, these methods can statistically infer whether a bifurcation is approaching \cite{Boettiger2012JRSocInterface, Hessler2022NewJPhys}, addressing the first two limitations. Binary classifiers trained with machine learning can realize this goal as well \cite{Huang2024NatMachineIntel}.
Second, some model-based \cite{Hessler2022NewJPhys, Zhang2022NatEcolEvol, Ditlevsen2023NatCommun, Grziwotz2023SciAdv}
and machine learning \cite{Kong2021PhysRevResearch, LiuZhang2024PhysRevX, Panahi2024PhysRevResearch} EWS methods estimate the value of $u$ at which the bifurcation occurs. However, the accuracy of the estimated bifurcation point in low-data settings typical of many experiments remains unclear in most of these methods. Although some model-based methods provided confidence intervals for the estimated bifurcation point in relatively low data settings, such model-based methods need to assume a particular bifurcation scenario, most notably the saddle-node bifurcation \cite{Ditlevsen2023NatCommun}.

\begin{figure}[t]
\centering
\includegraphics[width=0.6\textwidth]{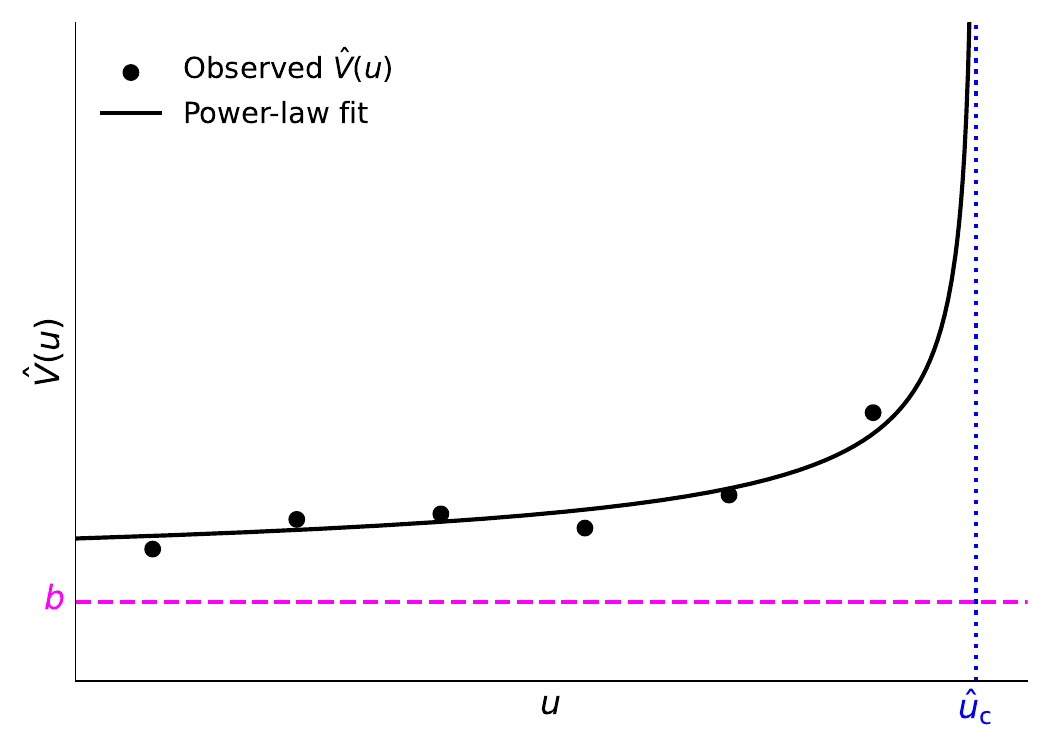}
\caption{Schematic of a power-law fit used by TIPMOC. The value of the control parameter at which the fitted power law diverges, $\hat{u}_{\text{c}}$, is the estimated bifurcation point.}
\label{fig:schem-powerlaw-fit}
\end{figure}

The goal of this study is to provide a method that, using only the sample variance (which is a widely used scalar EWS),
statistically tests whether a bifurcation is being approached and at which value of the control parameter, $u$, it will occur. We call this method TIPMOC (TIpping via Power-law fits and MOdel Comparison). TIPMOC monitors the sample variance as $u$ gradually changes, potentially toward a bifurcation. TIPMOC stops once it concludes that a bifurcation is imminent, outputting a future value of $u$ as predicted bifurcation point. Otherwise, TIPMOC terminates when it processes all available data, concluding that no impending bifurcation is detected. We do so by taking advantage of the mathematical property that some scalar EWSs, represented by the sample variance, diverge according to a particular power-law as a bifurcation point is approached. We fit such a power-law function to the observed pairs ($u$, $\hat{V}$), where $\hat{V}$ is the sample variance (see Fig.~\ref{fig:schem-powerlaw-fit} for a schematic) and compare a power-law divergence model versus alternative model, based on the ($u$, $\hat{V}$) relationship. Despite the simplicity of the approach, the method performs reasonably well for a range of dynamical systems and across a battery of stress tests, whereas our numerical results also show that the accuracy of estimating the bifurcation point is limited in many cases.

\section*{Results}

\subsection*{TIPMOC: Key idea and method overview\label{sec:theory}}

Consider a stochastic nonlinear dynamical system of any dimension. It may undergo a codimension-one bifurcation as a control parameter, $u$, monotonically changes and crosses the bifurcation point $u_{\text{c}}$. Parameter $u$ is often called the bifurcation parameter; we call it the control parameter to allow for cases where no bifurcation occurs. We assume that we observe $L$ samples of a scalar observable of the dynamical system, $x$, at stationarity at each value of $u$ and compute the EWS. We demonstrate our method for the sample variance of $x$, denoted by $\hat{V}$; the variance is a canonical scalar EWS motivated by critical slowing down~\cite{Carpenter2006EcolLett, Scheffer2009Nature, Dakos2015PhilTransRSocB, Southall2021JRSocInterface}. However, TIPMOC is also applicable to other scalar EWSs that show power-law divergence near $u_{\text{c}}$ (see the Discussion section for examples).

We assume that we observe a sequence of $\hat{V}$ as the control parameter monotonically changes, $\{\hat{V}(u_1), \hat{V}(u_2), \ldots \}$ (filled circles in Fig.~\ref{fig:schem-powerlaw-fit}). We develop a method that monitors sequentially arriving $(u_1, \hat{V}(u_1))$, $(u_2, \hat{V}(u_2))$, and so forth, and alerts the approach to the bifurcation point at the alarm point $u_{\ell}$. If TIPMOC detects an impending bifurcation, then we also provide $\hat{u}_{\text{c}}$ ($> u_{\ell}$), computed based on observations up to $u = u_{\ell}$, as the predicted bifurcation point. In this case, we then stop and do not use observations for $u_{\ell+1}$, $u_{\ell+2}$, $\ldots$. If no alert is triggered by the final $u$ value, TIPMOC concludes that there is no impending bifurcation.

We use the fact that true variance $V(u)$ shows power-law divergence as $u \to u_{\text{c}}$ for different types of codimension-one bifurcations \cite{Bury2020JRSocInterface}. To explain this point, we consider the normal forms of saddle-node, transcritical, pitchfork, and Hopf bifurcations \cite{Kuznetsov2000book, Strogatz2015book}. For these types of bifurcations, it is known for the true variance, $V(u)$, that $V(u) \propto 1/\left|\text{Re}(\lambda)\right|$ and that the bifurcation occurs when $\text{Re}(\lambda)$ crosses zero from negative to positive values, where $\lambda$ is the leading eigenvalue of the Jacobian matrix, i.e., the eigenvalue with the largest real part, and $\text{Re}(\lambda)$ represents the real part of $\lambda$. In the normal form of the saddle-node bifurcation, i.e., $\text{d}x/\text{d}t = u + x^2$, the bifurcation occurs at $u=0$ as $u$ increases from a negative value. For $u < 0$, we obtain $\lambda = 2x^* = - 2\sqrt{-u}$, where $x^*$ ($<0$) denotes the stable equilibrium. Therefore, $V(u)$ diverges according to $V(u) \propto (-u)^{-1/2}$ as $u \uparrow 0$. In the normal form of the transcritical bifurcation, i.e., $\text{d}x/\text{d}t = ux - x^2$, the bifurcation occurs at $u=0$. In this case, the stable equilibrium is $x^* = 0$ for $u<0$, and we obtain $\lambda = u-2x^*$. Therefore, $V(u) \propto (-u)^{-1}$ as $u \uparrow 0$. An analogous calculation shows that $V(u) \propto (-u)^{-1}$ for the normal forms of the supercritical and subcritical pitchfork bifurcations as $u \uparrow 0$, which is the bifurcation point. The same scaling law holds for the normal form of the Hopf bifurcation (see Methods for the derivation). The same power-law relationship between $u$ and $V(u)$ also holds for dynamical systems on networks, including for node-averaged variance~\cite{Masuda2024NatCommun}.

Therefore, we fit a power-law form
\begin{equation}
V(u) = a (\hat{u}_{\text{c}} - u)^{-\gamma} + b
\label{eq:power-law-fit}
\end{equation}
(shown as the solid curve in Fig.~\ref{fig:schem-powerlaw-fit}) to the observed sequence of EWS values, $\{\hat{V}(u_1), \ldots, \hat{V}(u_{\ell}) \}$ (shown as filled circles), and refit the model each time a new observation of $(u, \hat{V}(u))$ becomes available. We compute $\hat{V}(u)$ using $L=100$ samples of $x$ at a given value of $u$. It is straightforward to apply TIPMOC to the case in which $\hat{V}(u)$ is computed over a range of $u$ values using rolling time windows such that computation of $\hat{V}(u)$ at nearby $u$ values uses overlapping samples. For expository purposes, here we assume to observe $\hat{V}(u)$ for monotonically increasing values of $u$, whereas TIPMOC is equally applicable when $u$ monotonically decreases by changing Eq.~\eqref{eq:power-law-fit} to $V(u) = a (u - \hat{u}_{\text{c}})^{-\gamma} + b$, where $ u > \hat{u}_{\text{c}}$. Because we expect $V(u)$ to diverge in an inverse power of $u_{\text{c}} - u$ as $u \uparrow u_{\text{c}}$, we estimate the parameters, $a$, $\hat{u}_{\text{c}}$, $\gamma$, and $b$, subject to the constraints that $a>0$, $\hat{u}_{\text{c}} > \max \{ u_1, \ldots, u_{\ell} \}$, $\gamma > 0$, and $b < \min \{ \hat{V}(u_1), \ldots, \hat{V}(u_{\ell}) \}$; see Fig.~\ref{fig:schem-powerlaw-fit} for these constraints on $\hat{u}_{\text{c}}$ and $b$. We report $\hat{u}_{\text{c}}$ as our estimate of the bifurcation point. See Methods for details of the fitting procedure.

We fit Eq.~\eqref{eq:power-law-fit} to $\left\{ \left( u_1, \hat{V}(u_1) \right), \ldots, \left( u_{\ell}, \hat{V}(u_{\ell}) \right) \right\}$ as $\ell$ increases.
(We remind that each $\hat{V}(u_{\ell'})$ (with $1\le \ell' \le \ell$) is computed from $L$ samples of $x$ recorded at $u = u_{\ell'}$.)
Specifically, we first fit Eq.~\eqref{eq:power-law-fit} to $\left\{ \left( u_1, \hat{V}(u_1) \right), \ldots, \left( u_{\ell_0}, \hat{V}(u_{\ell_0}) \right) \right\}$, where $u_{\ell_0}$ is the last $u$ value in the initial fitting window, and $\ell_0$ is the initial number of observations of the control parameter used for fitting. We set $\ell_0 = 8$ in all the numerical simulations in the present article.
%
%
We then refit the power-law to $\left\{ \left( u_1, \hat{V}(u_1) \right), \ldots, \left( u_{\ell}, \hat{V}(u_{\ell}) \right) \right\}$ with $\ell = \ell_0 + 1$, $\ell_0 + 2$, $\ldots$. We repeat this procedure until we detect an impending bifurcation or reach the final available $u$ value.

At each $\ell$, we compute the goodness of fit of the power-law function, Eq.~\eqref{eq:power-law-fit}, based on the sum of squared errors of the fit. Then, we compare it to the goodness of fit for a linear fit (i.e., $V(u) = \alpha u + \beta$) to the same data, $\left\{ \left( u_1, \hat{V}(u_1) \right), \ldots, \left( u_{\ell}, \hat{V}(u_{\ell}) \right) \right\}$. We calculate the corrected Akaike Information Criterion, AIC$_{\text{c}}$~\cite{Burnham2002book}, which adjusts the AIC for small sample sizes (see Methods for the definition) for both power-law and linear fits. When the AIC$_{\text{c}}$ for the power-law fit is smaller than that for the linear fit by at least $10$ \cite{Burnham2002book, Symonds2011BehavEcolSociobiol} for three consecutive indices $\ell$, we declare an impending bifurcation. In this case, we output $\hat{u}_{\text{c}}$, and stop updating for further values of $u$. The requirement of three consecutive threshold crossings (i.e.,  AIC$_{\text{c}}$ difference of at least $10$) is a guard against isolated outliers \cite{Lehnertz2024Chaos}. TIPMOC is parametric and avoids explicit estimation of the underlying stochastic dynamical system generating the data. If the AIC$_{\text{c}}$ difference (linear minus power law) never exceeds 10 for any three consecutive values of $u_{\ell}$, then we say that we find no evidence of an approaching bifurcation.

\subsection*{Demonstration}

To demonstrate the method, we consider a stochastic double-well dynamical system \cite{Witt1997PhysRevE, Gammaitoni1998RevModPhys} governed by
\begin{equation}
\text{d}x = \left[ -(x - r_1)(x - r_2)(x - r_3) + u \right] \text{d}t + \sigma \text{d}W.
\label{eq:doublewell}
\end{equation}
Parameters $r_1$, $r_2$, $r_3$ are fixed constants satisfying $r_1 < r_2 < r_3$; $u$ is the control parameter; $\sigma$ represents the strength of dynamical noise; $W(t)$ is a Wiener process. We set $r_1 = 1$, $r_2 = 3$, $r_3 = 5$, and $\sigma = 0.25$. In the absence of dynamical noise (i.e., $\sigma = 0$), the dynamics has a unique stable equilibrium at $x^* < r_1$, which we call the lower equilibrium, when $u$ is sufficiently small. When $u$ is sufficiently large, the unique stable equilibrium is located at $x^* > r_3$, which we call the upper equilibrium. In an intermediate range of $u$, the lower equilibrium and the upper equilibrium are bistable. When $u$ gradually increases from a sufficiently small value, the double-well dynamics initialized near the lower equilibrium undergoes a saddle-node bifurcation when the lower equilibrium collides with an unstable equilibrium and disappears. This bifurcation point is located at $u \approx 3.079$
%
%
in the deterministic case. We aim to anticipate this bifurcation.

\begin{figure}[t]
\centering
\includegraphics[width=0.33\textwidth]{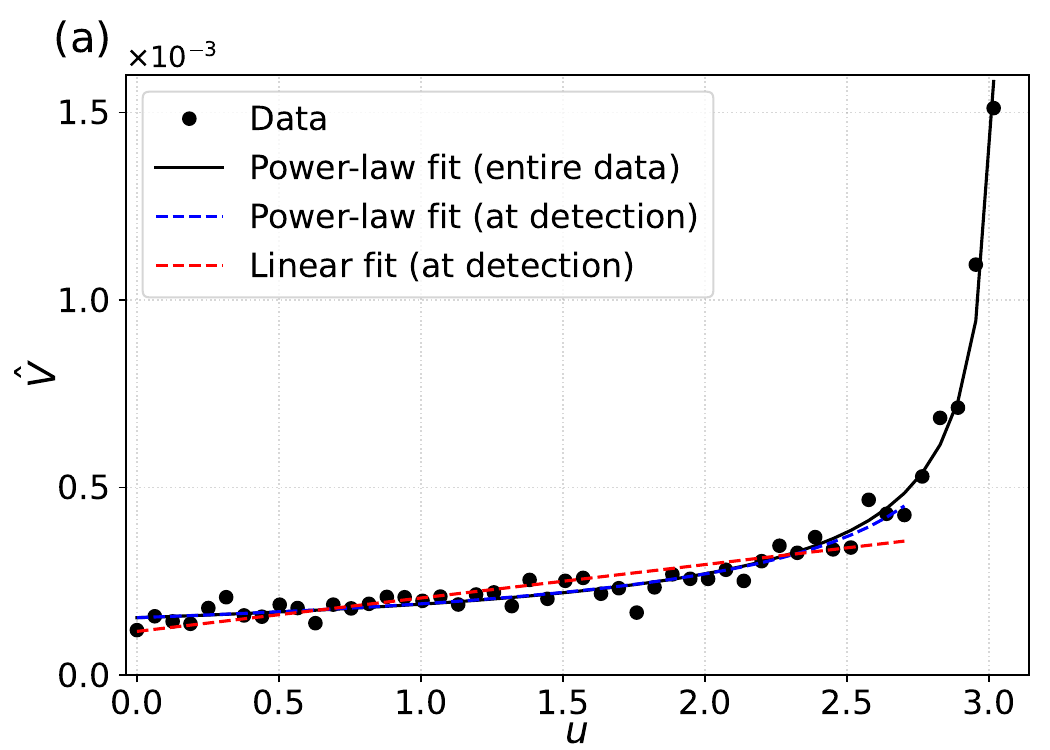}
\includegraphics[width=0.33\textwidth]{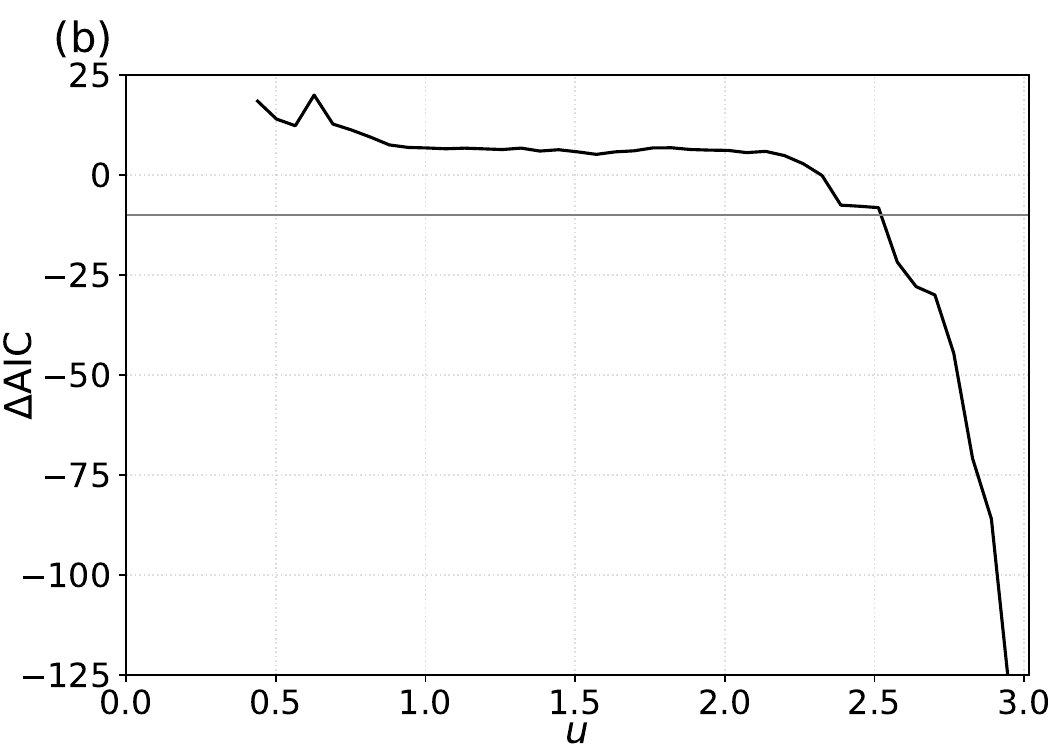}
\includegraphics[width=0.33\textwidth]{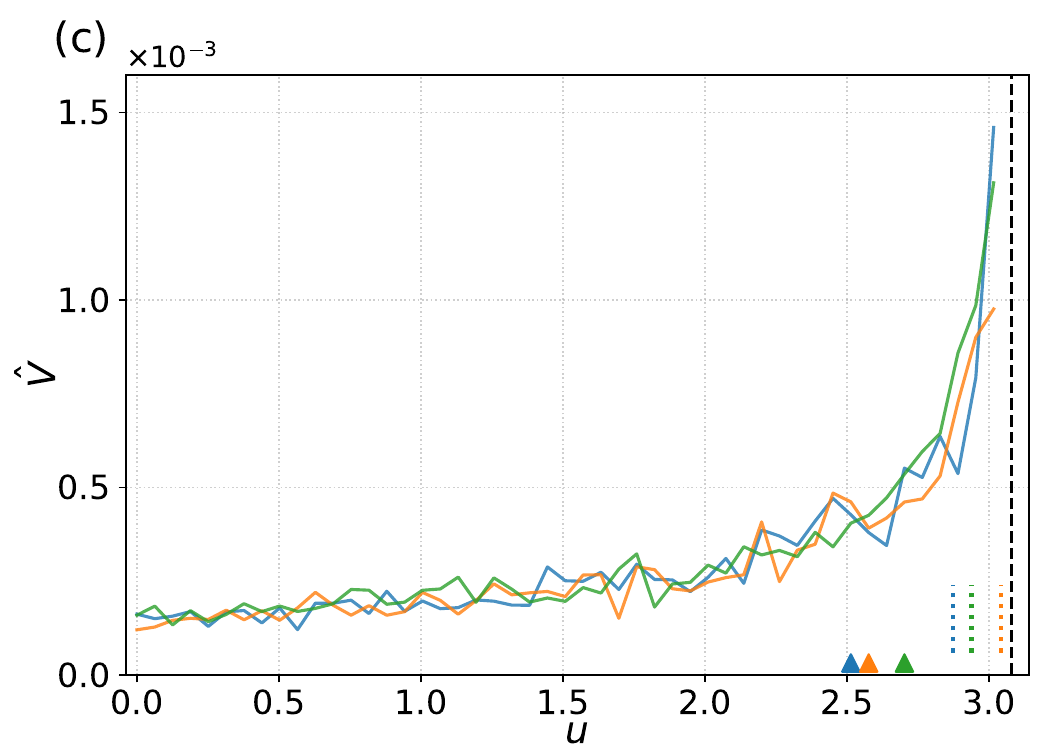}
\caption{Detection of the saddle-node bifurcation point for the stochastic double-well system. (a) $\hat{V}$ as a function of $u$, shown as circles, for one simulation. An impending bifurcation is detected at $u_{\text{det}} = 2.702$, and the bifurcation point is estimated as $\hat{u}_{\text{c}} = 3.185$. The solid line represents the power-law fit to all the observed ($u$, $\hat{V}$) pairs. The blue dashed line represents the power-law fit at detection (i.e., fit using the data up to $u_{\text{det}}$). The red dashed line represents the least-square linear fit at detection. (b) $\Delta \text{AIC}_{\text{c}}$ as a function of $u$. The horizontal line represents the detection threshold, $\Delta \text{AIC}_{\text{c}} = -10$. We do not show
$\Delta \text{AIC}_{\text{c}}$ for early $u$ values because we start the model fit and comparison at $\ell_0 = 8$ to avoid comparing between fits based on too few data points. (c) $\hat{V}$ as a function of $u$ for three simulation runs. Each color represents a simulation. The triangles and dotted lines represent $u_{\text{det}}$ and $\hat{u}_{\text{c}}$, respectively, with the same color convention. The dashed line represents the deterministic bifurcation point, $u = u_{\text{c}}$.} 
\label{fig:double-well-1-run}
\end{figure}

In Fig.~\ref{fig:double-well-1-run}(a) and (b), we show numerical results for one simulation run in which we measure $\hat{V}$ on an evenly spaced grid of $u$ values starting from $u=0$ and ending at the largest $u$ value before the saddle-node bifurcation. For this and all subsequent numerical simulations, we use at most $50$ monotonically increasing (or decreasing) values of $u$ before the bifurcation. Figure~\ref{fig:double-well-1-run}(a) shows that $\hat{V}$ increases towards the bifurcation point as $u$ increases, yielding Kendall's $\tau = 0.770$. While this $\tau$ value is substantially large, it does not distinguish between a linear trend and a superlinear trend of $\hat{V}(u)$ nor estimate the bifurcation location $u_{\text{c}}$. As theoretically expected, fitting Eq.~\eqref{eq:power-law-fit} to the full set of $\hat{V}$ values (including those closest to the bifurcation) yields a good match to the data (solid line in Fig.~\ref{fig:double-well-1-run}(a)).
 
For each $\ell \ge 8$, we show $\Delta \text{AIC}_{\text{c}}$ computed from $\{ (u_1, \hat{V}(u_1)), \ldots, (u_{\ell}, \hat{V}(u_{\ell}) \}$ in Fig.~\ref{fig:double-well-1-run}(b). The $u$ value on the horizontal axis of the figure represents $u_{\ell}$, the last $u$ value in the fitting window. We find that the linear fit is better than the power-law fit, producing positive $\Delta \text{AIC}_{\text{c}}$ values, when $u$ is approximately less than $2.5$. As $u$ increases further, $\Delta \text{AIC}_{\text{c}}$ becomes negative and then satisfies $\Delta \text{AIC}_{\text{c}} < -10$ for three consecutive $u$ values for the first time with the third consecutive crossing occurring at $u_{\text{det}} = 2.702$. (The subscript `det' denotes the detection point.) The power-law and linear fits using data up to $u_{\text{det}}$ are shown as the blue and red dashed lines, respectively, in Fig.~\ref{fig:double-well-1-run}(a). At $u = u_{\text{det}}$, we declare that a bifurcation will occur at some $u > u_{\text{det}}$. The fitted power law diverges at $\hat{u}_{\text{c}} = 3.185$, which is our estimate of the bifurcation point. This value is close to the actual bifurcation point in the deterministic case, $u \approx 3.079$. This is consistent with the similarity between the power-law fit based on $u \in [0, u_{\text{det}}]$ (blue dashed line) and the fit based on the full range of $u$ (black solid line). We note that, in the presence of dynamical noise, noise can induce a transition before the bifurcation point in the deterministic case \cite{Horsthemke1984book, Lindner2004PhysRep}. Therefore, one should not be overly optimistic about the accuracy of the estimated bifurcation point. The power-law fit at $u = u_{\text{det}}$ also returns $\gamma = 0.573$, which is not far from the theoretically expected value for the saddle-node bifurcation, $\gamma = 0.5$. 

In other simulations, $\hat{u}_{\text{c}}$ (and the estimated $\gamma$ value) may substantially deviate from the deterministic bifurcation point $u_{\text{c}}$. We show $\hat{V}$ as a function of $u$ for three simulation runs in Fig.~\ref{fig:double-well-1-run}(c). Each color corresponds to one simulation run. Both $u_{\text{det}}$ (shown by the triangles) and $\hat{u}_{\text{c}}$ (shown by the dotted lines) vary from run to run. While all three $\hat{u}_{\text{c}}$ values shown in Fig.~\ref{fig:double-well-1-run}(c) are smaller than $u_{\text{c}}$, this is not true in general. Nevertheless, when $\hat{V}$ has reached approximately $0.45$, TIPMOC correctly concludes that the bifurcation will occur (also see Fig.~\ref{fig:double-well-1-run}(a)). To further examine robustness of the results, we have carried out $100$ simulation runs, each starting with $u=0$ and increasing $u$ gradually. In all $100$ runs, TIPMOC has detected the impending saddle-node bifurcation point (i.e., $u \approx 3.079$ in the absence of dynamical noise) before the bifurcation point is reached (see the ``\% Detected'' column of Table~\ref{tab:detection-summary}).

\begin{figure}[t]
\centering
\includegraphics[width=0.49\textwidth]{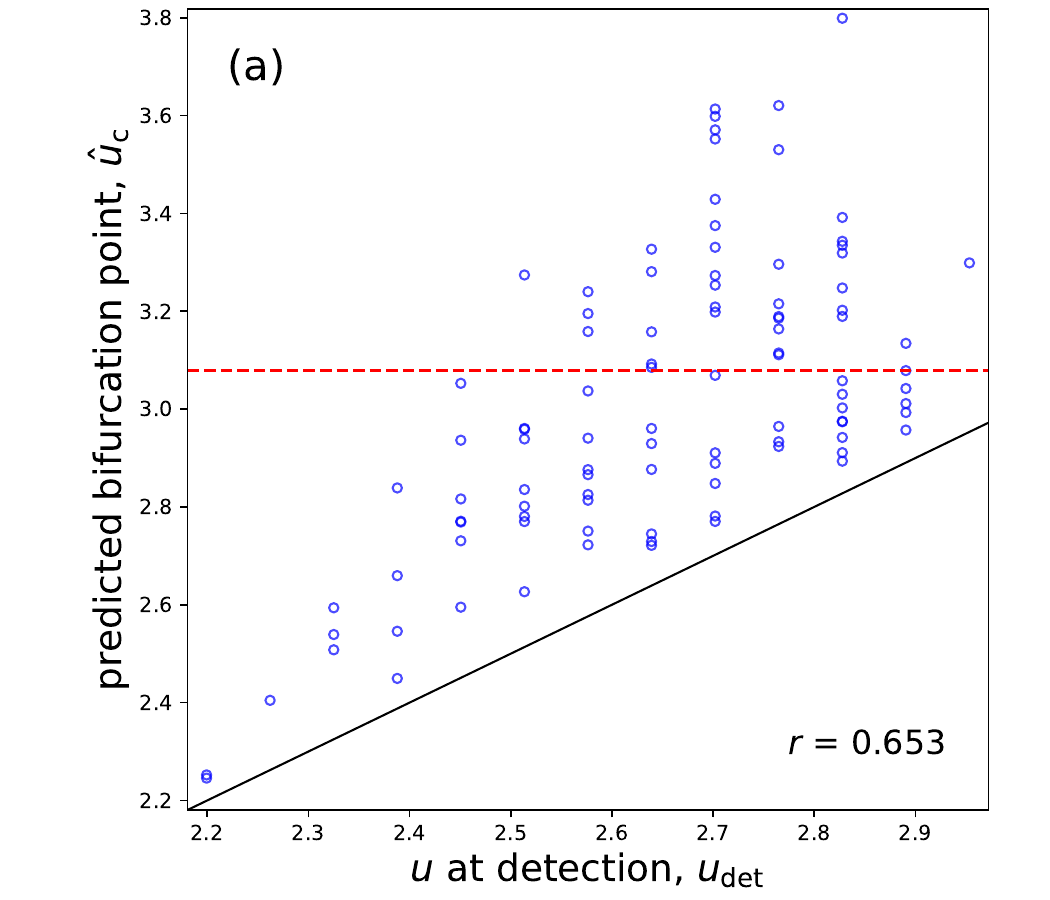}
\includegraphics[width=0.49\textwidth]{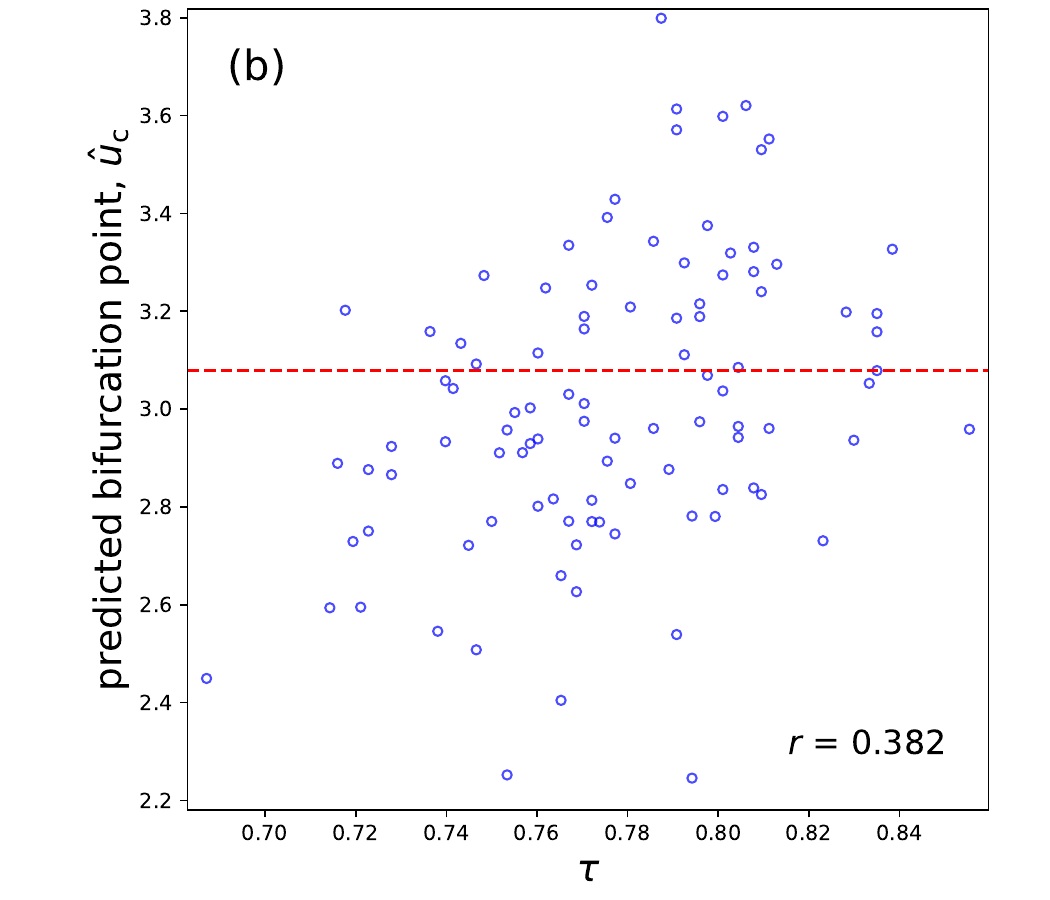}
\caption{Results of early detection of the bifurcation point for $100$ runs of the stochastic double-well dynamics.
(a) Relationship between the control parameter at detection, i.e., $u_{\text{det}}$, and the bifurcation point predicted at $u = u_{\text{det}}$, i.e., $\hat{u}_{\text{c}}$. (b) Relationship between Kendall's $\tau$ and $\hat{u}_{\text{c}}$. Each circle corresponds to one run. The horizontal dashed lines show the deterministic saddle-node bifurcation point, $u_{\text{c}} \approx 3.079$, in the absence of dynamical noise. In panel (a), the solid line represents the identity line $\hat{u}_{\text{c}} = u_{\text{det}}$. In (a) and (b) each, we also show the Pearson correlation coefficient between the two quantities across the $100$ runs.}
\label{fig:scatter_dw}
\end{figure}

We show the relationship between $u_{\text{det}}$ and $\hat{u}_{\text{c}}$ in Fig.~\ref{fig:scatter_dw}(a). Each circle in the figure corresponds to one run. The horizontal dashed line shows the saddle-node bifurcation point in the deterministic case (i.e., $\sigma = 0$). By construction, all the runs satisfy $\hat{u}_{\text{c}} > u_{\text{det}}$; the solid line represents the identity line $\hat{u}_{\text{c}} = u_{\text{det}}$ as a guide. The figure indicates that $u_{\text{det}}$ and $\hat{u}_{\text{c}}$ are positively correlated, with Pearson $r = 0.653$. This result implies that, if the bifurcation point is detected later, then the estimated location of the bifurcation point tends to be larger. Figure~\ref{fig:scatter_dw}(a) indicates that, when the impending bifurcation is detected at $u < 2.4$, the predicted position of the bifurcation, $\hat{u}_{\text{c}}$, is well below $3.079$. Therefore, an early detection, which is desirable, comes at a cost of reporting the bifurcation point far below the true value. When the impending bifurcation is detected at $2.4 < u < 3.079$, which accounts for 90\% of runs, $\hat{u}_{\text{c}}$ tends to be close to the deterministic value (i.e., $u_{\text{c}} \approx 3.079$), but with a substantial run-to-run variability.

To quantify the spread of $\hat{u}_{\text{c}}$, we show the mean $\pm$ standard deviation of $\hat{u}_{\text{c}}$ across 100 runs in Table~\ref{tab:detection-summary} (see the $\hat{u}_{\text{c}}$ column). We also report the fraction of runs satisfying $\hat{u}_{\text{c}} \in [u_{\text{c}} - \Delta u_{\text{c}}, u_{\text{c}} + \Delta u_{\text{c}}]$ in the same table. We set $\Delta u_{\text{c}} = 0.1(u_{\text{c}} - u_1)$; we recall that $u_1 = 0$ is the initial value of the control parameter. These statistics confirm that $\hat{u}_{\text{c}}$ is distributed reasonably concentrated around $u_{\text{c}}$ despite some outliers visible in Fig.~\ref{fig:scatter_dw}(a).

The mean of Kendall's $\tau$ across the $100$ runs is $0.777$ (see the $\tau$ column of Table~\ref{tab:detection-summary}). We also find that $\tau$ is not strongly correlated with $\hat{u}_{\text{c}}$ and that a large $\tau$ is not associated with an accurate $\hat{u}_{\text{c}}$ (see Fig.~\ref{fig:scatter_dw}(b)).

We conclude that TIPMOC successfully detects an impending saddle-node bifurcation for this stochastic dynamical system. TIPMOC also provides a point estimate of the bifurcation point, $\hat{u}_{\text{c}}$, at detection point $u_{\text{det}}$. However, $\hat{u}_{\text{c}}$ is only of moderate accuracy, which may be because we have only allowed at most $50$ pairs of $(u, \hat{V}(u))$.

\begin{table}
\centering
  \caption{Detection of bifurcations and their estimated locations for various stochastic dynamical systems. The numbers shown are based on $100$ runs of each system. All quantities except $\tau$ are computed only using the runs with successful detection. The fraction of runs with successful detection is shown in the ``\% Detected'' column. For the mutualistic-interaction dynamics, we show a sign-flipped Kendall's $\tau$ in the ``$\tau$'' column such that a large value corresponds to a good performance regardless of the dynamical system.}
  \label{tab:detection-summary}
  \begin{tabular}{lcccccc}
    \toprule
    Dynamics & {$\tau$} & {\% Detected} & $u_{\text{c}}$ & \multicolumn{2}{c}{$\hat{u}_{\text{c}}$} & $\text{Corr}(u_{\text{det}}, \hat{u}_{\text{c}})$ \\
    \midrule
    Double-well & $0.777 \pm 0.033$ & 100\% & 3.079 & $3.010 \pm 0.299$ & 70.0\% in & 0.653\\
    (equally spaced $u$, white noise)                    & & & & & $[2.771, 3.387]$ \\
    Double-well & $0.759 \pm 0.041$ & 93\% & 3.079 & $2.942 \pm 0.358$ & 59.1\% in & 0.690\\
    (random $u$, white noise)                    & & & & & $[2.771, 3.387]$ \\
    Double-well & $0.839 \pm 0.027$ & 99\% & 3.079 & $3.687 \pm 3.859$ & 32.3\% in & 0.281\\
    (random $u$, colored noise)                    & & & & & $[2.771, 3.387]$ \\
    Over-harvesting ($K=10$) & $0.680 \pm 0.039$ & 99\% & 2.604 & $2.548 \pm 0.085$ & 90.9\% in & 0.729\\
                      & & & & & $[2.444, 2.764]$ \\
    Linear grazing & $0.791 \pm 0.025$ & 96\% & 1 & $0.997 \pm 0.142$ & 59.4\% in & 0.565\\
                      & & & & & $[0.900, 1.100]$ \\
    Rosenzweig-MacArthur & $0.741 \pm 0.041$ & 100\% & 2.6 & $2.379 \pm 0.118$ & 21.0\% in & 0.661\\
                      & & & & & $[2.450, 2.750]$ \\
    Mutualistic interaction & $0.805 \pm 0.031$ & 99\% & 0.047 & $-0.032 \pm 0.195$ & 54.5\% in & 0.313\\
                      & & & & & $[-0.048, 0.142]$ \\
    OU &  $0.873 \pm 0.019$ & 0\% & N/A & N/A & N/A & N/A \\
    Over-harvesting ($K=2$) & $0.425 \pm 0.060$ & 0\% & N/A & N/A & N/A & N/A \\
    \bottomrule
  \end{tabular}
\end{table}

\subsection*{Robustness tests}

We perform the following robustness tests for TIPMOC.

\subsubsection*{Other dynamics}

We apply TIPMOC to other nonlinear dynamical systems showing bifurcations that are commonly used in EWS and theoretical-ecology literature. We use a one-dimensional over-harvesting model of ecological dynamics with $K=10$ (see Methods for the definition of $K$), which shows a saddle-node bifurcation \cite{Noymeir1975JEcol, May1977Nature, Guttal2008EcolLett, Scheffer2009Nature, Dakos2012PlosOne, Guttal2013TheorEcol, Kefi2013Oikos, Dutta2018Oikos, Bury2021PNAS}. We also use a one-dimensional over-harvesting model with a linear grazing functional response showing a transcritical bifurcation \cite{Kefi2013Oikos}. We additionally use a two-dimensional consumer-resource model (Rosenzweig-MacArthur model with a type II functional response) showing a Hopf bifurcation \cite{Rosenzweig1963AmNat, Rosenzweig1971Science, Kefi2013Oikos}. Finally, we use a mutualistic-interaction model on a $100$-node synthetic network that shows mass extinction \cite{GaoBarzelBarabasi2016Nature}. We run this model until the first node shifts from the upper to lower state. In this model, we gradually decrease the strength of interspecific coupling to probe the onset of mass extinction of species as environment conditions deteriorate. In the three over-harvesting models, we gradually increase the control parameter as in Ref.~\cite{Kefi2013Oikos}.

\begin{figure}[t]
\centering
\includegraphics[width=0.48\textwidth]{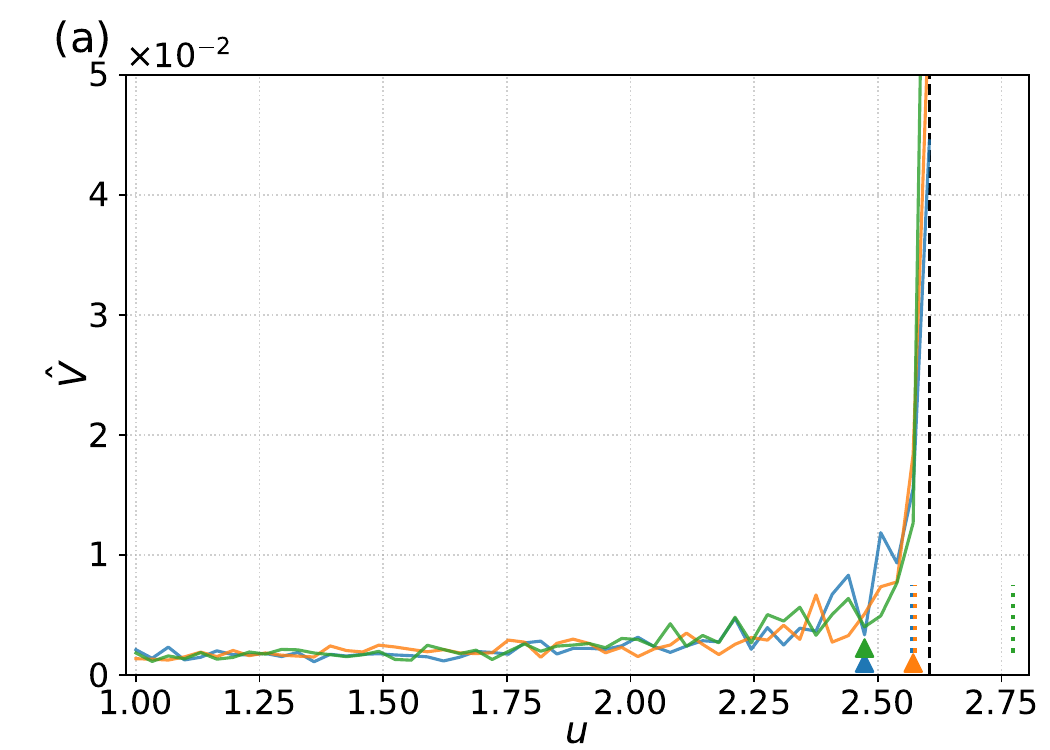}
\includegraphics[width=0.48\textwidth]{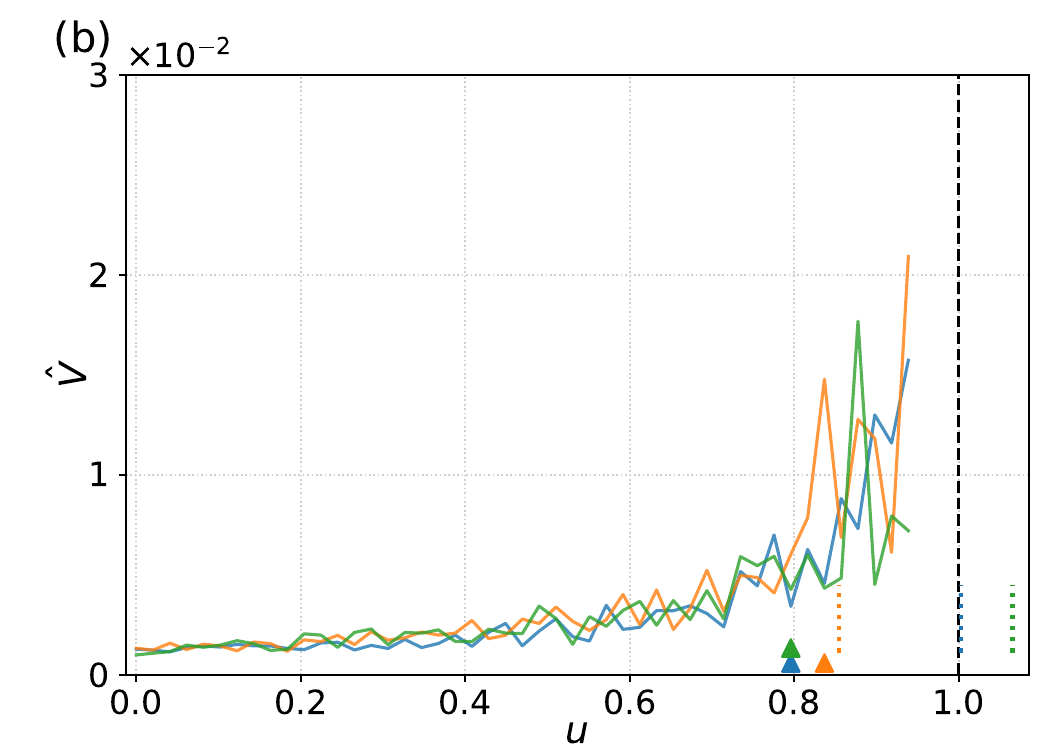}
\includegraphics[width=0.48\textwidth]{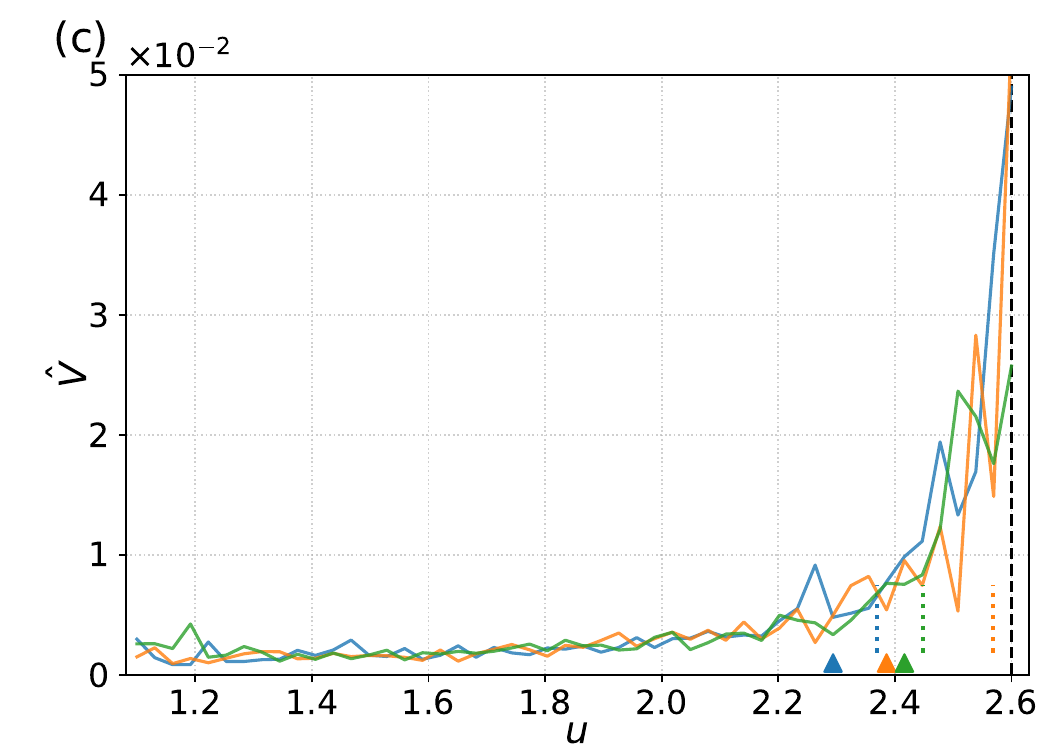}
\includegraphics[width=0.48\textwidth]{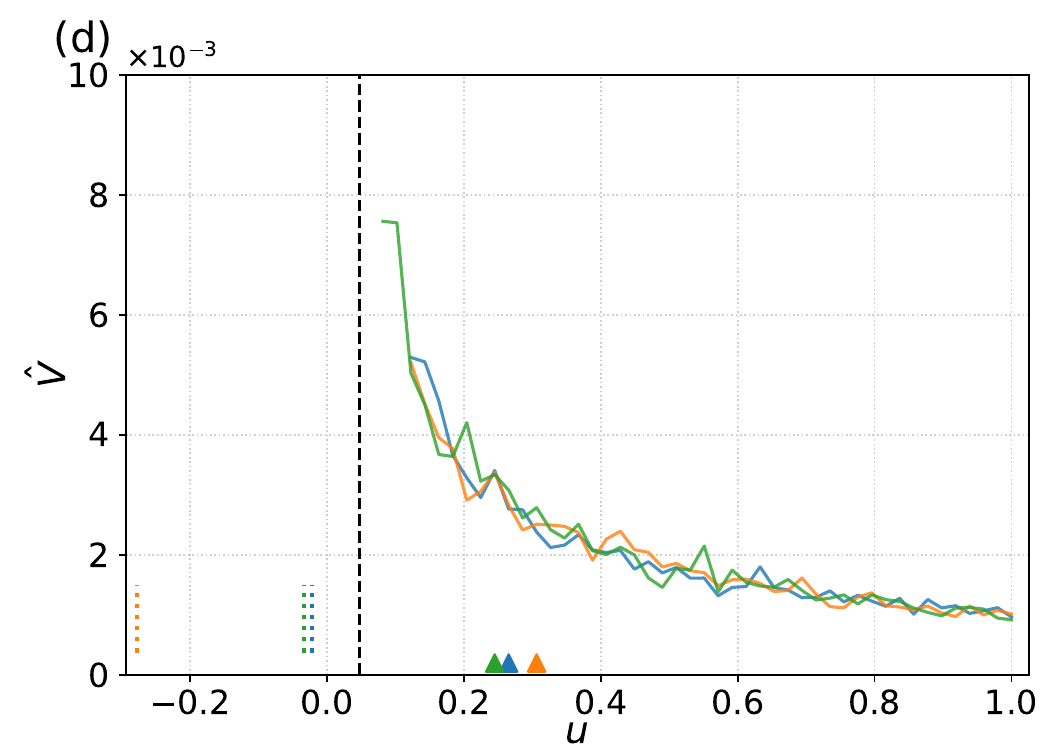}
\caption{$\hat{V}$ for three runs of different dynamical systems and detection of impending bifurcations. (a) Over-harvesting model with $K=10$, showing a saddle-node bifurcation. (b) Over-harvesting model with linear grazing, showing a transcritical bifurcation. (c) Rosenzweig-MacArthur model, showing a Hopf bifurcation. (d) Mutualistic-interaction model on a synthetic network with $100$ nodes having community structure. Each color denotes one run. Each panel has three runs. The triangles and dotted lines represent $u_{\text{det}}$ and $\hat{u}_{\text{c}}$, respectively. The dashed lines represent the bifurcation point in the deterministic case, $u_{\text{c}}$. In all runs shown in this figure, TIPMOC detects an impending bifurcation before reaching the deterministic bifurcation point. In (d), we have gradually decreased the control parameter, $u$, representing the coupling strength among different species, to emulate deterioration toward mass extinction. See Methods for the network used in (d).} 
\label{fig:3-runs-for-various-dynamics}
\end{figure}

In Fig.~\ref{fig:3-runs-for-various-dynamics}, we show $\hat{V}$ as a function of $u$ for three simulation runs for each dynamical system. Note that we have stopped the runs earlier than the deterministic bifurcation point (the dashed lines) in Fig.~\ref{fig:3-runs-for-various-dynamics}(b) and (d) because dynamical noise induced transitions earlier. The figure indicates that, in all runs and dynamical systems, TIPMOC detects the impending stochastic bifurcation event before it occurs. This result is encouraging given that $\hat{V}$ substantially varies across runs, in particular in the over-harvesting models (Fig.~\ref{fig:3-runs-for-various-dynamics}(a), (b), and (c)). The figure also indicates that the estimated bifurcation point, $\hat{u}_{\text{c}}$, varies widely.

We have further run $100$ simulations for each dynamical system. TIPMOC detects the impending bifurcation event before reaching the deterministic bifurcation point (i.e., $u_{\text{c}}$) in all runs or in the vast majority of runs for each system, as shown in Table~\ref{tab:detection-summary}. The table also shows that the detection results are qualitatively similar to those for the double-well system in two respects. First, $\hat{u}_{\text{c}}$ is positively correlated with the $u$ value at detection (i.e., $u_{\text{det}}$). Therefore, an early detection of bifurcation tends to reduce the accuracy of $\hat{u}_{\text{c}}$. Second, $\hat{u}_{\text{c}}$ is centered near $u_{\text{c}}$ with substantial dispersion across runs. In fact, we observe differences in the accuracy of $\hat{u}_{\text{c}}$ across dynamical systems, particular in the fraction of runs yielding $\hat{u}_{\text{c}} \in [u_{\text{c}} - \Delta u_{\text{c}}, u_{\text{c}} + \Delta u_{\text{c}}]$. This fraction ranges between 21\% and 91\% depending on the dynamical system. However, we should interpret this result with caution because
the fraction depends on our operational tolerance band $[u_{\text{c}} - \Delta u_{\text{c}}, u_{\text{c}} + \Delta u_{\text{c}}]$. The band width,
$\Delta u_{\text{c}}$, depends on the initial value of $u$, i.e., $u_1$, which is our arbitrary choice. Therefore, we conclude that TIPMOC is robust across dynamical systems showing different types of bifurcations in that it successfully detects the impending bifurcation with a high detection rate. We also conclude that $\hat{u}_{\text{c}}$ is close on average to $u_{\text{c}}$, but with substantial variability across runs.

\subsubsection*{Uneven spacing of the control parameter}

In practice, the values of the control parameter at which we collect EWS observations may be unknown. In particular, observations may occur at unevenly spaced $u$ values \cite{Zhuge2025RSOpenSci}, including the case in which $u$ evolves nonlinearly in time \cite{Obrien2023NatCommun}. Because TIPMOC requires paired observations $\{ (u_1, \hat{V}(u_1)), (u_2, \hat{V}(u_2)), \ldots \}$, if we only observe a sequence of $\hat{V}$ values without knowing the corresponding $u$ values, TIPMOC is not directly applicable. In this situation, we propose to use TIPMOC by assuming that $u$ is evenly distributed. To test the robustness of TIPMOC in this scenario, we simulated the stochastic double-well dynamical system, Eq.~\eqref{eq:doublewell}, with exponentially distributed spacing between two adjacent $u$ values (i.e., $u_{i+1} - u_i$). Then, we pretend that we do not know the $u$ values at which we measure $\hat{V}$, presume that they are equidistantly distributed (although this is false), and apply TIPMOC. The detection results are shown in Table~\ref{tab:detection-summary} (see the ``random $u$, white noise'' row). In this challenging setting, we find that TIPMOC detects the impending saddle-node bifurcation in most of runs and that the accuracy of $\hat{u}_{\text{c}}$ remains comparable to when $u$ is known.

\subsubsection*{Colored noise}

So far, we have assumed that the dynamical noise is white. Performance of EWSs based on critical slowing down may degrade if the dynamical noise is not white \cite{Perretti2012EcolAppl, Dutta2018Oikos, QinTang2018PhysRevE, Hessler2022PnasNexus, Kuehn2022ProcRSocA, Proverbio2023Iscience, Morr2024PhysRevResearch}. Therefore, here we investigate the double-well model with colored noise. We find that the probability that the impending bifurcation is detected remains high even when $u$ is unevenly spaced and the noise is colored simultaneously (see the ``random $u$, colored noise'' row in Table~\ref{tab:detection-summary}). This result is consistent with prior results that the sample variance is relatively strong against colored noise compared to other EWSs \cite{Perretti2012EcolAppl, Hessler2022PnasNexus}.
However, the table also shows that colored noise severely degrades the accuracy of $\hat{u}_{\text{c}}$.

\subsubsection*{False positives}

An important requirement for EWSs is a low false-positive rate, i.e., not to detect an impending bifurcation when no bifurcation is expected \cite{Boettiger2013TheorEcol, Dablander2023PsycholMethods}. Therefore, we simulate one-dimensional Ornstein-Uhlenbeck processes in which the variance of $x$ increases linearly with $u$. It should be noted that no bifurcation occurs in this model. TIPMOC has flagged an upcoming bifurcation event in none of the $100$ runs (see Table~\ref{tab:detection-summary}). The table also shows that $\tau$ is comparably large to that in the bifurcating dynamical systems examined above. We also examine the over-harvesting model that we already used for a saddle-node bifurcation, but now with a different parameter value (i.e., $K=2$) for which varying the control parameter does not produce a bifurcation \cite{Kefi2013Oikos}. TIPMOC does not detect an upcoming bifurcation in any of the $100$ runs for this model of dynamics either (see Table~\ref{tab:detection-summary}). Therefore, we conclude that TIPMOC can avoid false positives.

\section*{Discussion\label{sec:discussion}}

We have exploited the observation that the sample variance diverges in a particular power-law form as a bifurcation point is approached. Thus, we have proposed TIPMOC. TIPMOC alerts an impending bifurcation in a statistically controlled manner and forecasts the location of the bifurcation. We have shown that TIPMOC detects the impending bifurcation with a low false negative rate. It also avoids false positives when the dynamical system gradually changes but does not approach a bifurcation. Avoiding false positives of this type may be difficult for methods that aim to detect a statistically significant deviation of a scalar EWS or estimated model from a baseline \cite{Ditlevsen2023NatCommun, Proverbio2023Iscience}. TIPMOC is robust across dynamical systems, uneven spacing of sampled $u$ values, and colored noise. TIPMOC is equally applicable to other scalar EWSs that diverge according to a power-law. Such EWSs include the standard deviation (i.e., $\sqrt{\hat{V}(u)}$) \cite{Dakos2012PlosOne, Kefi2013Oikos, Southall2021JRSocInterface}, sample variance (or standard deviation) averaged over nodes in the given network \cite{Maclaren2023JRSocInterface, Masuda2024NatCommun, Yu2026PhilTransRSocA}, the sample variance of the node-averaged state \cite{Patterson2021AmNat, Morr2024SiamJApplDynSyst}, and the leading eigenvalue of the covariance matrix \cite{Brock2006EcolSoc, Dakos2018EcolInd, ChenOdea2019SciRep}. It should be noted that TIPMOC is not applicable to the lagged autocorrelation, which is another widely used scalar EWS \cite{Dakos2012PlosOne, Dakos2015PhilTransRSocB, Southall2021JRSocInterface}. This is because lagged autocorrelation is bounded in $[-1, 1]$ and therefore does not show divergence as the tipping point is approached. In contrast, TIPMOC is intended for variance-based EWSs.

Apart from machine learning approaches, which we will discuss below, there are other EWS methods that aim to infer whether a bifurcation is imminent, or forecast the location of the bifurcation point. These approaches use explicit stochastic dynamical modeling of the data-generating process. They rely on model comparison \cite{Boettiger2012JRSocInterface, Hessler2022NewJPhys, Hessler2022PnasNexus, Hessler2025NatCommun}, model fitting but without model comparison \cite{Ditlevsen2023NatCommun}, or on estimating the Jacobian matrix of the dynamics \cite{Grziwotz2023SciAdv}. Although a seminal study on EWSs using statistical model comparison works with a small amount of data \cite{Boettiger2012JRSocInterface}, it seems that testing whether a bifurcation is being approached and when requires a relatively large $L$, i.e., the number of samples per control parameter value or per environmental condition. For example, reported sample sizes include $L=1000$ \cite{Hessler2022NewJPhys}, $L=750$ \cite{Hessler2022PnasNexus} (but they also showed their method often works well with $L=100$ and $L=150$), $L=2000$ \cite{Hessler2025NatCommun}, and $L=250$ \cite{Grziwotz2023SciAdv}. TIPMOC avoids fitting an explicit dynamical system model to the observed data. Instead, it carries out model comparison at the level of a summary statistic, $\hat{V}$. While scalar EWSs using temporal statistics typically require 
$L$ that may exceed what is available in many experiments or applications (e.g., $L=100$, as used in the present study), TIPMOC may be relatively data efficient among EWSs that do not use spatial (i.e., multivariate) statistics. 
On the other hand, it is possible to apply TIPMOC without any modification to a sequence of sample variances computed with sliding windows (i.e., to compute $\hat{V}$ from samples of $x$ that fall in a prescribed time window, and one slides such time windows allowing overlaps between them). This approach is common in the analysis of experimental data. It is data efficient although it sacrifices the sensitivity of $\hat{V}$ to changes in the dynamical system or environment due to a low-pass filtering nature of sliding window analysis.
We leave a systematic comparison of data efficiency for future work.

There is increasing use of machine learning for anticipating tipping events \cite{Kong2021PhysRevResearch, Patel2023Chaos, Huang2024NatMachineIntel, Panahi2024PhysRevResearch, LiuZhang2024PhysRevX} and for classifying bifurcation types \cite{Bury2021PNAS, Bury2023NatCommun, Deb2024RSocOpenSci}.
We do not claim that TIPMOC outperforms any of these methods. Rather, TIPMOC is transparent, built directly on conventional scalar EWSs, and can be layered onto studies that already compute $\hat{V}$. While TIPMOC is complementary to these machine learning approaches, TIPMOC can be readily combined with machine learning to potentially improve detection of bifurcation and/or estimation of $u_{\text{c}}$. This is because TIPMOC takes as input only sequential pairs $\{ (u_1, \hat{V}(u_1)), (u_2, \hat{V}(u_2)), \ldots, \}$. A machine-learning variant of TIPMOC would train a neural network (or a different architecture) on various data sequences $\{ (u_1, \hat{V}(u_1)), (u_2, \hat{V}(u_2)), \ldots, \}$ generated from various bifurcating and non-bifurcating dynamical systems. Because TIPMOC uses $\hat{V}$ sequences rather than raw state trajectories, the most suitable architectures of neural networks may differ from those used for early warning from raw time series data. We defer this topic to future work.

Performance in estimating the bifurcation point varied across different dynamical systems (see Table~\ref{tab:detection-summary}). For the discussion,
we assume that $u$ gradually increases. In general, the accuracy would depend on the range of the control parameter we examine, $[u_1, u_{\text{c}}]$, where we recall that $u_1$ is the initial value of the control parameter. For a fixed absolute error (i.e., $\left| \hat{u}_{\text{c}} - u_{\text{c}} \right|$), a relative error,
 $\left| \hat{u}_{\text{c}} - u_{\text{c}} \right| / (u_{\text{c}} - u_1)$, decreases when $u_1$ becomes small relative to $u_{\text{c}}$. Note that we often cannot interpret the absolute error because we may not know the scale of $u$ in many cases. For example, we cannot tell whether $\left| \hat{u}_{\text{c}} - u_{\text{c}} \right| = 0.1$ is small or large unless we know the physical meaning of $u$ or the rate at which $u$ changes. Because the relative error depends on $u_1$, we should take the values shown in Table~\ref{tab:detection-summary} (i.e., mean $\pm$ standard deviation of $\hat{u}_{\text{c}}$ and the within-band fraction of runs) only as a guide. For the same reason, we should not use these summaries to directly compare the performance of TIPMOC across different dynamical systems. Despite this caveat, the accuracy of TIPMOC at locating $u_{\text{c}}$ is not high overall and likely reflects the inherent difficulty in predicting $u_{\text{c}}$ \cite{Hessler2022NewJPhys}. This difficulty is shared by some other modern methods \cite{Hessler2022NewJPhys, Dablander2023PsycholMethods, Grziwotz2023SciAdv}, while higher accuracy at predicting $u_{\text{c}}$ has been reported in some cases \cite{Zhang2022NatEcolEvol, Ditlevsen2023NatCommun, LiuZhang2024PhysRevX}.
%
%
Improving the accuracy of estimating $u_{\text{c}}$ warrants future work. Possible strategies include improving the optimization procedure to fit the power-law curve, smoothing of $\hat{V}(u)$ before the fitting, using rolling windows to compute $\hat{V}$, and incorporating machine learning approaches.

We demonstrated that TIPMOC is capable of anticipating three common codimension-one bifurcations, i.e., saddle-node, transcritical, and Hopf bifurcations, using over-harvesting models \cite{Kefi2013Oikos}. This advantage contrasts with a recently proposed powerful model-based method that assumes saddle-node bifurcations~\cite{Ditlevsen2023NatCommun}. Some EWS methods and related approaches classify bifurcation types \cite{Bury2020JRSocInterface, Bury2021PNAS, Grziwotz2023SciAdv}. TIPMOC may partially support this task because $\hat{V}$ diverges as $(u_{\text{c}} - u)^{-1/2}$ as $u \uparrow u_{\text{c}}$ for the saddle-node bifurcation and as $(u_{\text{c}} - u)^{-1}$ for the transcritical, Hopf, and also pitchfork bifurcations \cite{Bury2020JRSocInterface, Kuehn2011PhysicaD}. Our method provides an estimate of the power-law exponent, $\gamma$, for $\hat{V} \propto (u_{\text{c}} - u)^{-\gamma}$. In fact, the estimated $\gamma$ value is highly variable in our current implementation of TIPMOC, likely due in part to using at most $50$ noisy ($u$, $\hat{V}(u)$) pairs. Therefore, we did not address the problem of classifying bifurcation types using TIPMOC in this study.

We did not apply TIPMOC to empirical data. This is because, although EWSs are theoretically grounded on critical slowing down, experimental evidence for divergence of variance near tipping points remains limited \cite{Dakos2015PhilTransRSocB, Obrien2023NatCommun, Helmich2024NatRevPsychol, Rietkerk2025NatClimateChange}. To the best of our knowledge, successful applications of the sample variance use a small number of parameter points \cite{ChenJayaprakash2018AmNat} or identify a tipping point because of a sustained elevation in sample variance rather than a clear divergence \cite{Carpenter2011Science, Wichers2016PsychotherapyPsychosomatics, Ditlevsen2023NatCommun}. High-frequency measurements such as remote sensing techniques \cite{Dakos2015PhilTransRSocB, Delecroix2023PlosGlobalPublicHealth} or innovative sampling designs \cite{Helmich2024NatRevPsychol} may provide sufficiently dense data to test for scaling consistent with variance divergence. 
We also remark that we have focused on the case of effectively one-dimensional critical dynamics in this article. Intricate types of dynamical noise such as time-dependent \cite{Morr2024PhysRevResearch}, multiplicative \cite{Morr2024PhysRevResearch}, or common noise applied across different variables \cite{Morr2024SiamJApplDynSyst} can degrade performance of variance-based (or other) EWSs in multi-dimensional dynamical systems. In addition, colored and other types of non-Markovian dynamical noise often degrade performance of EWSs based on critical slowing down
 \cite{Perretti2012EcolAppl, Dutta2018Oikos, QinTang2018PhysRevE, Hessler2022PnasNexus, Kuehn2022ProcRSocA, Proverbio2023Iscience}. Our numerical results on compromised performances in estimating the bifurcation point under colored noise compared to under white noise are consistent with these previous findings. Addressing these issues is important for applying EWS methods, including TIPMOC, to empirical data.
At present, TIPMOC may find more applications in analyzing dynamical-system models. 
A comprehensive evaluation of TIPMOC on various tipping events and continuous-onset bifurcations, including higher-dimensional models and a wider variety of dynamical noise, is an immediate direction of future work. Furthermore, comparison across modern EWS methods including those based on model inference or machine learning is largely lacking. Future studies should carry out such head-to-head comparisons as well.

\section*{Methods\label{sec:Methods}}

\subsection*{Sample variance for the normal form of a Hopf bifurcation}

The normal form of the supercritical Hopf bifurcation is given in Cartesian coordinates by
\begin{equation}
\begin{cases}
\dot{x} &= ux - y + (\alpha x - \beta y) \left( x^2 + y^2\right),\\
\dot{y} &= x + uy + (\beta x + \alpha y) \left( x^2 + y^2\right),
\end{cases}
\label{eq:Hopf-normal}
\end{equation}
where $u$ is the control parameter, and $\alpha$ and $\beta$ are constants.
The dynamical system given by Eq.~\eqref{eq:Hopf-normal} has $(x, y) = (0, 0)$ as the unique stable equilibrium when $u < 0$. This equilibrium undergoes a Hopf bifurcation at $u=0$ to become unstable as $u$ increases through zero. The Jacobian of Eq.~\eqref{eq:Hopf-normal} at $(x, y) = (0, 0)$ has eigenvalues $\lambda = u \pm i$.

By linearizing Eq.~\eqref{eq:Hopf-normal} about $(x, y) = (0, 0)$ and adding dynamical noise independent in $x$ and $y$, we obtain
\begin{equation}
\begin{cases}
\text{d}x &= (ux - y) \text{d}t + \sigma \text{d}W_1,\\
\text{d}y &= (x + uy) \text{d}t + \sigma \text{d}W_2,
\end{cases}
\label{eq:Hopf-linear}
\end{equation}
where $W_1(t)$ and $W_2(t)$ are independent Wiener processes. The Lyapunov equation~\cite{Gajic1995book,Gardiner2009book} for Eq.~\eqref{eq:Hopf-linear} is given by
\begin{equation}
\begin{pmatrix} -u & 1 \\ -1 & -u \end{pmatrix}
C + C
\begin{pmatrix} -u & -1 \\ 1 & -u \end{pmatrix}
=
\begin{pmatrix} \sigma^2 & 0 \\ 0 & \sigma^2 \end{pmatrix},
\label{eq:Lyapunov-Hopf}
\end{equation}
where $C \in \mathbb{R}^{2\times 2}$ is the covariance matrix.
The solution of Eq.~\eqref{eq:Lyapunov-Hopf} is given by
\begin{equation}
C = -\frac{\sigma^2}{2u}
\begin{pmatrix} 1 & 0 \\ 0 & 1 \end{pmatrix}.
\end{equation}
Therefore, the variances of $x$ and $y$ are both equal to $-\sigma^2/(2u)$ ($>0$); note that $u<0$ before the Hopf bifurcation.

\subsection*{Models of dynamics\label{sub:dynamics}}

We used the following nonlinear dynamical systems commonly used in EWS research.

The stochastic double-well dynamical system is given by Eq.~\eqref{eq:doublewell}. 
At each value of $u$, we initialize each simulation run at $x = 1$ (i.e., $x = r_1$) such that the trajectory remains near the lower stable equilibrium when it is stable. We gradually increase $u$ from $0$ up to $3.079$ until a tipping occurs. Note that, in the absence of dynamical noise (i.e., $\sigma = 0$), a saddle-node bifurcation occurs at $u \approx 3.079$.

In the case of exponentially distributed intervals between consecutive $u$ values, we drew $50$ values of $u$ uniformly at random on $[0, 3.079]$ and sorted them in ascending order. This procedure creates a sequence of $u$ whose positions approximately obey a Poisson process. In other words, the interval between the adjacent $u$ values approximately obeys an exponential distribution.

We generated colored dynamical noise by an OU process
$\text{d}\xi = -\frac{\xi}{t_0} \text{d}t + \sigma \text{d}W$ with $t_0 = 1$. We initialize the process with $\xi(t=0) = 0$. This OU process only generates the dynamical noise, $\xi(t) \text{d}t$, to be added to $x$ in the stochastic double-well system in place of the white noise term, $\sigma \text{d}W$, in Eq.~\eqref{eq:doublewell}. This colored-noise OU process is distinct from the OU process used for false-positive tests (shown in Eq.~\eqref{eq:OU} below).

A model of over-harvesting is given by
\begin{equation}
\text{d}x = \left[ r x \left(1-\frac{x}{K} \right) - \frac{u x^2}{x^2 + \tilde{x}^2} \right] \text{d}t + \sigma \text{d}W,
\label{eq:harvesting}
\end{equation}
where $x$ is the resource biomass, $r$ is the growth rate, $K$ is the carrying capacity, $u$ is the maximum grazing rate, and $\tilde{x}$ is a parameter adjusting the grazing rate \cite{Noymeir1975JEcol, May1977Nature, Scheffer2009Nature, Kefi2013Oikos}. We use $u$ as the control parameter. We set $r=1$, $K=10$, and $\tilde{x} = 1$ following previous literature \cite{Guttal2008EcolLett, Dakos2012PlosOne, Guttal2013TheorEcol, Kefi2013Oikos, Dutta2018Oikos}. 
We also set $\sigma = 0.05$. While multiplicative dynamical noise can be used for this model \cite{Dakos2012PlosOne, Dutta2018Oikos}, here we use additive dynamical noise, as in Eq.~\eqref{eq:harvesting}; this is a common choice \cite{Guttal2008EcolLett, Dakos2010TheorEcol, Guttal2013TheorEcol, Kefi2013Oikos, Bury2021PNAS}.
%
We initialize each run at $x = K$. We gradually increase $u$ from $1$ up to $2.604$ until a tipping occurs. In the absence of dynamical noise, a saddle-node bifurcation occurs at $u = u_{\text{c}} \approx 2.604$ \cite{Dakos2012PlosOne}, and $x$ collapses to a value near zero for $u > u_{\text{c}}$.

When we examine a scenario in which no tipping occurs, we set $K=2$, for which $x$ responds smoothly to changes in $u$, without a bifurcation \cite{Kefi2013Oikos}.

The over-harvesting model with a linear grazing functional response~\cite{Kefi2013Oikos} is given by
\begin{equation}
\text{d}x = \left[ r x \left(1-\frac{x}{K} \right) - u x \right] \text{d}t + \sigma \text{d}W.
\label{eq:linear-grazing}
\end{equation}
We set $r=1$ and $K=10$ following previous literature \cite{Kefi2013Oikos}. We also set $\sigma = 0.05$. At each $u$, We initialize at the corresponding nontrivial equilibrium, $x = (r-u)K/r$. We gradually increase $u$ in $[0, 1]$. Note that the transcritical bifurcation occurs at $u=r=1$ in the absence of dynamical noise.

A model of consumer-resource dynamics, which we refer to as the Rosenzweig-MacArthur model with type II functional response \cite{Rosenzweig1963AmNat, Rosenzweig1971Science, Kefi2013Oikos}, is given by
\begin{align}
\text{d}x &= \left[ r x \left(1-\frac{x}{K} \right) - \frac{gxy}{x+h} \right] \text{d}t + \sigma \text{d}W_1,\\
\text{d}y &= \left[ \frac{egxy}{x+h} - my \right] \text{d}t + \sigma \text{d}W_2,
\end{align}
where $x$ is the biomass of the resource, and $y$ is the biomass of the consumer. We set the growth rate of the resource $r=0.5$, maximum grazing rate $g=0.4$, half-saturation constant $h=0.6$, assimilation efficiency of consumer $e=0.6$, and mortality rate of consumer $m=0.15$ \cite{Kefi2013Oikos}. The dynamical noise added to $x$ and $y$, i.e., Wiener processes $W_1(t)$ and $W_2(t)$, are assumed to be independent. For simplicity, we set $\sigma=0.01$ in both equations. The nontrivial equilibrium for which a Hopf bifurcation occurs in the absence of noise, $(x^*, y^*)$, is given by $x^* = mh/(eg-m)$ and $y^* = r(x^*+h)/g \times (1-x^*/K)$. We use the carrying capacity $K$ as the control parameter (i.e., $K = u$) \cite{Kefi2013Oikos}. At each $K$, we initialize each run at $(x, y) = (x^*, y^*)$. Feasibility of the nontrivial equilibrium, i.e., $x^*, y^* > 0$, requires $eg-m > 0$, which is satisfied with our parameter values, and $K > x^* = 1$. The Hopf bifurcation of equilibrium $(x^*, y^*)$ occurs at $K = h(eg+m)/(eg-m) = 2.6$ as $K$ gradually increases. Therefore, we gradually increase $K$ in $[1.1, 2.6]$. We exclude $[1, 1.1]$ because, in numerical simulations, $y < 0$ is frequently observed when $K \in [1, 1.1]$ due to dynamical noise. For $K \in [1.1, 2.6]$, we force $x = 0$ and $y = 0$ whenever we obtain $x<0$ and $y<0$, respectively, in the numerical simulation.

The stochastic mutualistic-interaction dynamics among species is given by
\begin{equation}
\text{d}x_i = \left[B_i + x_i \left( 1 - \frac{x_i}{K_i} \right) \left( \frac{x_i}{C_i} - 1 \right) + \tilde{u} +
D \sum_{j=1}^N w_{ij} \frac{x_i x_j}{\tilde{D}_i + E_i x_i  + H_j x_j} \right] \text{d}t + \sigma \text{d}W_i,
\label{eq:mutualistic}
\end{equation}
where $x_i$ represents the biomass of the $i$th species; $N$ is the number of nodes in the network; $B_i$, $C_i$, $\tilde{D}_i$, $E_i$, $H_i$, $K_i$, and $\tilde{u}$ (with $i\in \{1, \ldots, N\}$) are constants \cite{GaoBarzelBarabasi2016Nature}. Constant $B_i$ represents the migration rate of the $i$th species from outside the ecosystem; $C_i$ is the Allee constant; $K_i$ is the carrying capacity; $D$ is the coupling strength between adjacent nodes; $w_{ij}$ ($\ge 0$) represents the strength of the mutualistic effect of the $j$th species on the $i$th species. We set $B_i = 0.1$, $C_i = 1$, $\tilde{D}_i = 5$, $E_i = 0.9$, $H_i = 0.1$, and $K_i = 5$, $\forall i\in \{1, \ldots, N\}$, following previous literature~\cite{GaoBarzelBarabasi2016Nature, Masuda2024NatCommun}, and $\tilde{u} = - 0.29$ and $\sigma = 0.15$. We use $D$ as the control parameter (i.e., $u = D$). With these parameter values, in the deterministic case, we have numerically found that the tipping of the first node from the upper state (i.e., $x_i$ being near $K_i$) to the lower state (i.e., $x_i$ being near 0) occurs at $D \approx 0.047$ as $D$ gradually decreases from a larger value. While tipping of other nodes occurs as $D$ is decreased further from $D \approx 0.047$, we attempt to detect the first tipping at $D \approx 0.047$ using the EWS. At each $D$, we initialize the run at $x_i = 5$, $\forall i$ \cite{Masuda2024NatCommun}. With this initial condition, all the nodes are near their upper state, corresponding to the species' persistence, for each value of $D$ that sustains the upper equilibrium. We gradually decrease $D$ from $1$ toward $0.047$ until the first tipping occurs.

We use a network generated by the Lancichinetti-Fortunato-Radicchi (LFR) model, which produces networks with a heterogeneous degree distribution and communities of heterogeneous sizes \cite{Lancichinetti2008PhysRevE}. We use the same network instance (and the corresponding network-generation parameter values) as that used in our previous study \cite{Maclaren2025NatCommun}. The network is undirected and unweighted. Therefore, we set $w_{ij} = w_{ji} = 1$ if the $i$th and $j$th nodes are adjacent. Otherwise, we set $w_{ij} = w_{ji} = 0$.

The OU process is given by
\begin{equation}
\text{d}x = -\frac{x}{u} \text{d}t + \sigma \text{d}W,
\label{eq:OU}
\end{equation}
where $u > 0$. This model does not show bifurcations as $u$ varies. We set $\sigma=0.1$. We initialize each run at the equilibrium, $x=0$. We increase $u$ over $[0.01, 2]$. The stationary variance, which is the expectation of $\hat{V}$, is given by $\sigma^2 u/2$. Therefore, we expect $\hat{V}$ to increase linearly with $u$.

\subsection*{Simulation methods and computation of early warning signals\label{sub:EWS}}

We use the Euler-Maruyama method with a time step of $\Delta t=10^{-3}$ to simulate each stochastic dynamical system.

For each control parameter value held fixed, we run the simulation and compute the EWS as follows. We discard the first $10$ time units (TUs) of each run as transients before sampling to compute the EWS. We then collect $L=100$ samples of $x$ by recording $x$ every $T_{\text{skip}}$ TUs. The first sampling occurs at time $T_{\text{skip}}$ after transients are discarded. We set $T_{\text{skip}} = 1$ for the double-well model, over-harvesting models with nonlinear and linear grazing, and mutualistic-interaction model, and $T_{\text{skip}} = 10$ for the Rosenzweig-MacArthur model and OU process. This choice reduces serial correlation such that adjacent samples are approximately uncorrelated. For the Rosenzweig-MacArthur model, we sample values from $x$ (as opposed to $y$). For the mutualistic-interaction model, we sample values from $x_1$, i.e., the abundance of the first species, which is an arbitrary choice. Then, we compute $\hat{V}$ as the unbiased variance (i.e., sample variance with denominator $L-1$) of the $L=100$ samples.

For each dynamical system, we gradually change the control parameter (denoted by $u$ generically). We use 50 values of $u$ that are equally spaced in the specified range except when we examine randomly distributed $u$ values with the double-well model. For bifurcating models, we cap the parameter sweep at the bifurcation point in the deterministic case (i.e., the first bifurcation point when $\sigma = 0$). We declare a tipping event (i.e., a large regime shift) when $x$ is no longer near its initial state for the first time in the case of dynamical systems showing a saddle-node bifurcation. This case includes the double-well model, over-harvesting model with nonlinear grazing with $K=10$, and mutualistic-interaction model. We say that $x$ is no longer near its initial state if $x > r_2 = 3$ for the double-well model, $x < 0$ for the over-harvesting model, and $\min_{i \in \{1, \ldots, N\}} x_i < 0.1$ for the mutualistic-interaction model. If a tipping event is declared, then we stop the parameter sweep. Due to the dynamical noise, the sweep tends to stop at a parameter value
closer to start (i.e., smaller $u$ for the double-well and over-harvesting models and larger $u$ for the mutualistic-interaction model) than in the deterministic case. The value of $u$ at which the sweep stops may differ from run to run.

\section*{Kendall's $\tau$}

As a benchmark, we compute Kendall's $\tau$ from pairs $\{ \left( u_i, \hat{V}(u_i) \right) \}$. Note that $\tau$ lies in the range $[-1, 1]$. When the control parameter gradually increases (i.e., all but the mutualistic-interaction model), a large $\tau$ is associated with a good performance of the EWS. When the control parameter gradually decreases (i.e., the mutualistic-interaction dynamics), a strongly negative $\tau$ (i.e., close to $-1$) is better. Therefore, in Table~\ref{tab:detection-summary}, we show a sign-adjusted rank coefficient, $-\tau$, for the mutualistic-interaction dynamics, such that a large value is always desired \cite{Maclaren2025JRSocInterface}.

\section*{Fitting of a power-law function to EWS samples}

We fit Eq.~\eqref{eq:power-law-fit} to a given sequence of the sample variance, $\left\{ \left( u_1, \hat{V}(u_1) \right), \ldots, \left( u_{\ell}, \hat{V}(u_{\ell}) \right) \right\}$, via approximate minimization of the sum of squared errors as follows. We first tried the scipy.optimize.curve\_fit on SciPy (v1.17.0) / Python (v3.14.2) to estimate the four parameters, $a$, $\hat{u}_{\text{c}}$, $\gamma$, and $b$. However, presumably because this minimization is a highly non-convex problem, it did not converge in most cases. The fit tended to be poor even when the optimizer converged. Fixing $\gamma$ to the theoretically expected values for the codimension-one bifurcations (i.e., $\gamma = 0.5$ or $1$) and then estimating the other three parameters by curve\_fit still yielded many runs that did not converge. Therefore, we transform Eq.~\eqref{eq:power-law-fit} to
\begin{equation}
\ln \left( \hat{V} -b \right) = \ln a - \gamma \ln \left( \hat{u}_{\text{c}} - u \right)
\label{eq:power-law-fit-log}
\end{equation}
and run a linear fit between $\ln\left( \hat{u}_{\text{c}} - u \right)$ and $\ln \left( \hat{V} - b \right)$ for various choices of $\hat{u}_{\text{c}}$ and $b$ values. 

We assume that $u_1 < \cdots < u_{\ell}$. We optimize over the $(\hat{u}_{\text{c}}, b)$ parameter space specified by $u_{\ell} + \epsilon \le \hat{u}_{\text{c}} \le u_{\ell} + 10 (u_{\ell} - u_1)$ and $ \hat{V}_{\min} - \left( \hat{V}_{\max} - \hat{V}_{\min}\right)/2 \le b \le \hat{V}_{\min} - \epsilon$, where $\hat{V}_{\min} = \min \{ \hat{V}(u_1), \ldots, \hat{V}(u_{\ell}) \}$, $\hat{V}_{\max} = \max \{ \hat{V}(u_1), \ldots, \hat{V}(u_{\ell}) \}$, and $\epsilon = 10^{-5}$. This choice allows us to scan a wide range of possible bifurcation points (i.e., $\hat{u}_{\text{c}}$) relative to the range of the control parameter observed so far (i.e., $[u_1, u_{\ell}]$). We obtain the optimal $(\hat{u}_{\text{c}}, b)$ by minimizing the Pearson correlation coefficient (i.e., making it as close to $-1$ as possible) between $\ln \left( \hat{u}_{\text{c}} - u \right)$ and $\ln \left( \hat{V}(u) - b \right)$ across the $\ell$ data points. This is because we expect that $\gamma > 0$, such that Eq.~\eqref{eq:power-law-fit-log} implies that $\ln \left( \hat{u}_{\text{c}} - u \right)$ and $\ln \left( \hat{V}(u) - b \right)$ should be negatively correlated, and a Pearson correlation coefficient closer to $-1$ is better. We use scipy.optimize.differential\_evolution on SciPy (v1.17.0) for the optimization. We verified that the optimized Pearson correlation coefficient upon detection of an impending bifurcation was between $-1$ and $-0.8$ in most of the runs used in Table~\ref{tab:detection-summary}. Finally, for the optimized $\hat{u}_{\text{c}}$ and $b$ values, we obtained $\ln a$ (and hence $a$) and $\gamma$ by linear regression between $\ln \left( \hat{u}_{\text{c}} - u \right)$ and $\ln \left( \hat{V}(u) - b \right)$ given by Eq.~\eqref{eq:power-law-fit-log}. 

\section*{AIC$_{\text{c}}$}

We use AIC$_{\text{c}}$, which is a modification of the Akaike Information Criterion (AIC) for small sample sizes \cite{Burnham2002book}, to determine whether the linear or power-law model fits better to the data. Although the two models are not nested, one can use the AIC in general \cite{Burnham2002book}.
The AIC$_{\text{c}}$ is defined by
\begin{equation}
\text{AIC}_{\text{c}} = - 2 \ln \hat{\mathcal{L}} + 2k + \frac{2k(k+1)}{n-k-1},
\label{eq:AIC_c}
\end{equation}
where $\hat{\mathcal{L}}$ is the maximized likelihood for the model, $k$ is the number of parameters in the model, and $n$ is the number of observations. In our case, $n=\ell$ for $\{ (u_1, \hat{V}(u_1)), \ldots, (u_{\ell}, \hat{V}(u_{\ell})) \}$. The first two terms of Eq.~\eqref{eq:AIC_c} give the unmodified AIC. 

When $n$ is small, AIC$_{\text{c}}$ penalizes additional parameters more strongly than the unmodified AIC does. Therefore, model selection using AIC$_{\text{c}}$ favors models with fewer parameters when $n$ is small, compared to the unmodified AIC. A common rule of thumb is to use AIC$_{\text{c}}$ when $n/k < 40$ \cite{Burnham2002book}. In our case, the linear and power-law fits have $k=2$ and $k=4$, respectively. The number of data points, $n$, is at most $50$ in our analysis, justifying the use of AIC$_{\text{c}}$. We consider that using AIC$_{\text{c}}$ rather than the unmodified AIC is also appropriate in a variety of applications of EWSs. This is because at most tens, but not hundreds or thousands, of approximately independent $\left( u, \hat{V}(u)\right)$ samples may be observed in many situations.

Under the assumption of i.i.d.\,normal distribution of the residual, $-2 \ln \hat{\mathcal{L}}$ is equal to $n \ln (\text{RSS}/n)$, where $\text{RSS}$ is the sum of squared residuals across the data points, plus a constant that only depends on $n$ \cite{Burnham2002book, Symonds2011BehavEcolSociobiol}. We use this relationship to compute AIC$_{\text{c}}$ for both linear and power-law fits. It should be noted that our heuristic optimization procedure, explained in the last section, does not exactly maximize the likelihood. However, we compute AIC$_{\text{c}}$ for the optimized parameters of the power-law fit as an approximation.

We monitor $\Delta \text{AIC}_{\text{c}}$, which is AIC$_{\text{c}}$ for the linear model minus AIC$_{\text{c}}$ for the power-law model. A negative $\Delta \text{AIC}_{\text{c}}$ provides support for the power-law model over the linear model. We declare that the power-law model fits better than the linear model at the first $\ell$ at which the third consecutive threshold crossing occurs (i.e., $\Delta \text{AIC}_{\text{c}} < -10$ for $u_{\ell-2}$, $u_{\ell-1}$, and $u_{\ell}$). The threshold of $-10$ is a conservative choice \cite{Burnham2002book, Symonds2011BehavEcolSociobiol}.

\section*{Data availability}

The numerical data that are generated in this study are available on Github at \href{https://github.com/naokimas/tipmoc}{\textcolor{blue}{https://github.com/naokimas/tipmoc}}.

\section*{Code availability}

The code for generating the results and figures in this article is publicly available on Github at \href{https://github.com/naokimas/tipmoc}{\textcolor{blue}{https://github.com/naokimas/tipmoc}}.

\section*{Acknowledgements}

N.M. acknowledges support from the Japan Science and Technology Agency (JST) Moonshot R\&D (under grant no.\,JPMJMS2021), the National Science Foundation (under grant no.\,2204936), and JSPS KAKENHI (under grant nos.\,JP 23H03414, 24K14840, and 24K03013).



\section*{Competing interests}

The author declares no competing interests.

\end{document}